\documentclass[conference]{IEEEtran}
\IEEEoverridecommandlockouts
\usepackage{cite}
\usepackage{multirow}
\usepackage{booktabs}
\usepackage{amsmath,amssymb,amsfonts}
\usepackage{algorithmic}
\usepackage{graphicx}
\usepackage{textcomp}
\usepackage{graphicx}
\usepackage{hyperref}
\usepackage[table,xcdraw]{xcolor}

\def\BibTeX{{\rm B\kern-.05em{\sc i\kern-.025em b}\kern-.08em
    T\kern-.1667em\lower.7ex\hbox{E}\kern-.125emX}}
\begin{document}

\title{\fontsize{13}{16}\selectfont \textbf{CHALLENGING DATASET AND MULTI-MODAL GATED MIXTURE OF EXPERTS MODEL FOR REMOTE SENSING COPY-MOVE FORGERY UNDERSTANDING}\vspace{-10pt}}

\author{
    \IEEEauthorblockN{ Ze Zhang\textsuperscript{\dag},  Enyuan Zhao\textsuperscript{\dag}, Yi Jiang, Jie, Nie\textsuperscript{*} and Xinyue Liang}
    \IEEEauthorblockA{
        Faculty of Information Science and Engineering, Ocean University of China, Qingdao, China \\
        \{shenyedepisa, zhaoenyuan, jiangyi7601\}@stu.ouc.edu.cn, \{niejie, liangxinyue\}@ouc.edu.cn 
    \vspace{-10pt}}

\thanks{This work was supported by the National Natural Science Foundation of China (U22A2068, U23A20320 and 62302470) and the Fundamental Research Funds for the Central Universities(202313037). \vspace{-8pt}}

\thanks{

{\dag}: These authors have contributed equally to this work and share first authorship.

}
}
\maketitle
    
\begin{abstract}
The Remote Sensing Copy-Move Question Answering (RSCMQA) task focuses on interpreting complex tampering scenarios and inferring the relationships between objects. Currently, publicly available datasets often use randomly generated tampered images, which lack spatial logic and do not meet the practical needs of defense security and land resource monitoring. To address this, we propose a high-quality manually annotated RSCMQA dataset, Real-RSCM, which provides more realistic evaluation metrics for the identification and understanding of remote sensing image tampering. The tampered images in the Real-RSCM dataset are subtle, authentic, and challenging, posing significant difficulties for model discrimination capabilities. To overcome these challenges, we introduce a multimodal gated mixture of experts model (CM-MMoE), which guides multi-expert models to discern tampered information in images through multi-level visual semantics and textual joint modeling. Extensive experiments demonstrate that CM-MMoE provides a stronger benchmark for the RSCMQA task compared to general VQA and CMQA models. Our dataset and code are available at \href{https://github.com/shenyedepisa/CM-MMoE}{https://github.com/shenyedepisa/CM-MMoE}.
\end{abstract}


\section{Introduction}
\label{sec:intro}
Detecting forgery in remote sensing images is of paramount importance for land resource monitoring and national defense security, particularly for situational awareness during wartime. Copy-move image forgery, where a region of an image (source region) is copied to another location within the same image (tampered region), misleads users by hiding or highlighting specific objects, potentially causing economic losses and security risks. Since the tampered and source regions originate from the same image, their optical characteristics are almost identical. 
Additionally, the unique perspectives of remote sensing images, coupled with extensive monitoring areas and numerous small-sized targets, contribute to their semantic complexity and redundancy. This significantly complicates the detection of forgeries in remote sensing images.

Traditional single-modal copy-move forgery detection methods, which rely on visual features, primarily encompass feature-matching algorithms\cite{5438805,singh2022copy,gan2022novel,kaushal2024orbcmfd} and deep learning-based approaches\cite{xiong2023cmcf,weng2023ucm,wang2024object,islam2020doa,7123632}. However, the information extraction capabilities of single-modal models are inherently limited. To achieve precise detection and a deeper understanding of tampered images, current research paradigms are transitioning from single-modal to multimodal approaches. Recent research\cite{xu2024fakeshield} underscores that multimodal methods can significantly enhance forgery detection accuracy and provide rich semantic information that is directly interpretable by humans.


\begin{figure}[!t]
    \centering
    \includegraphics[width=0.98\linewidth]{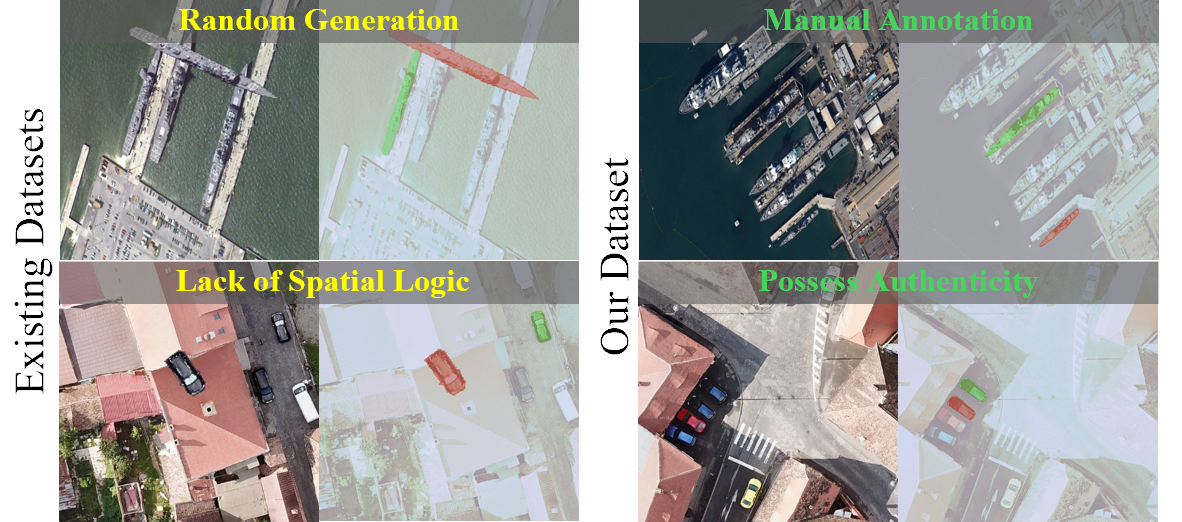}
    \vspace{-3pt}
    \caption{Examples of tampered images from existing datasets and our dataset. Randomly generated tampered images are easily identifiable and lack realism.
    }
    \vspace{-5pt}
    \label{fig:intro}
\end{figure}

In prior research, Lobry et al.\cite{Sylvain2020RSVQA} introduced the application of Visual Question Answering (VQA) techniques to extract critical information from remote sensing images. CDVQA\cite{yuan2022change} attempted to detect regional changes in remote sensing images of the same location at different times using VQA methods.
Zhang et al.\cite{zhang2024copymove} proposed a customized STMA method to interpret tampered scenes in remote sensing images. They developed a dataset of randomly generated remote sensing tampered images and trained a question-answering model based on multimodal injection of source and tampered region information, offering a preliminary solution for understanding copy-move forgery in remote sensing images. 
However, the current methods encounter several challenges: 
(1) The publicly available datasets utilize randomly generated tampered images, resulting in numerous images lacking spatial logic, such as vehicles appearing on rooftops and ships on land, as shown in Figure \ref{fig:intro}. Models trained on such data struggle to be applicable in real-world scenarios. 
(2) More subtle and realistic tampering poses significant challenges to the model's discrimination ability. In our previous research, the question-answering accuracy of existing models significantly decreased on high-quality tampered images.
To address these issues, we have made the following contributions:

\begin{figure*}[!t]
    \centering
    \includegraphics[width=0.99\textwidth]{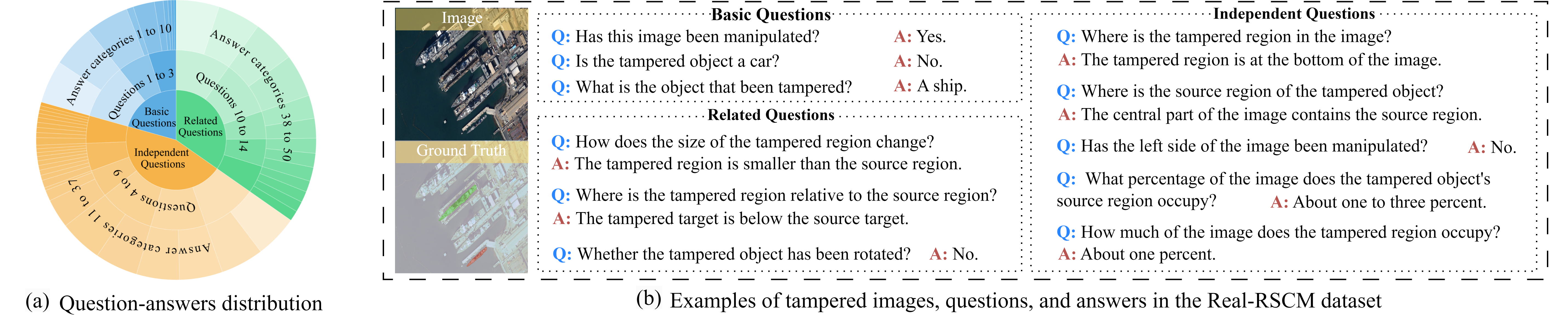}
    \vspace{-3pt}
    \caption{
    Distribution of questions and answers in the Real-RSCM dataset, along with specific examples. Basic questions assess overall image perception, independent questions target specific attributes of the source or tampered regions, and related questions extract the relative relationship between these regions.
    }
    \label{fig:dataset}
    \vspace{-5pt}
\end{figure*}

\begin{itemize}
  \item We introduce Real-RSCM, a high-quality, manually annotated RSCMQA dataset comprising 10k images and 173k image-question-answer triplets. The tampered locations in these images are subtle, realistic, and logically coherent.

  \item We propose a multimodal gated mixture of experts (CM-MMoE) model  for the RSCMQA task. By integrating various visual features with textual features, this model enables multi-expert systems to achieve a multi-level understanding of image semantics and accurately answer different types of questions.
  
  \item The CM-MMoE model is capable of extracting critical information from complex and redundant remote sensing tampering scenarios. Extensive comparative experiments and detailed ablation studies have demonstrated the effectiveness of our proposed method.

\end{itemize}

\section{Related Work}
Traditional copy-move forgery detection (CMFD) algorithms rely on block matching and keypoint matching techniques, such as PCA\cite{5438805}, DWT\cite{singh2022copy}, SIFT\cite{gan2022novel}, and ORB\cite{kaushal2024orbcmfd}. These methods depend heavily on strict prior information about image attributes, such as edge sharpness and local features, and can only perform mechanical feature matching without extracting semantic information. Given the exponential increase in image data, manually designing prior information for images is impractical. Consequently, deep learning-based methods have become the mainstream approach in recent years. BusterNet\cite{wu2018busternet} and STD-Net\cite{chen2020serial} use parallel and serial deep neural networks to extract source and tampered regions, respectively. Islam et al.\cite{islam2020doa} introduced Generative Adversarial Networks into CMFD task to improve localization accuracy. IMNet\cite{wang2024object} and CMCF-Net\cite{xiong2023cmcf} enhanced the model's receptive field by extracting coarse and fine features and stacking multi-scale features. TSCM-Net\cite{liu2021two} and UCM-Net\cite{weng2023ucm} utilized candidate feature proposal and instance segmentation to provide auxiliary information for tampering identification. Xu et al.\cite{xu2024fakeshield} latest research introduced multimodal models into the CMFD task, with the FakeShield model guiding the identification of tampered regions through question-driven mechanisms, providing rich explanatory information.

Due to the complexity of remote sensing images, Lobry et al.\cite{Sylvain2020RSVQA} proposed using VQA methods to extract key information. Siebert et al.\cite{Siebert2022MultiModal} employed VisualBERT to better learn joint representations. EarthVQA\cite{wang2024earthvqa} utilized segmentation masks to guide models in focusing on important information within the images. CDVQA\cite{yuan2022change} was the first to notice the issue of change detection in remote sensing images, attempting to perceive changes in the same location across different temporal phases. In the latest research, Zhang et al.\cite{zhang2024copymove} focused on solving the CMFD task in remote sensing images. They inject the source and tampered region information into the CMQA model to extract specific information about tampering. However, this model's feature extraction method is limited and has not been tested on high-difficulty tampered data. Moreover, there are deficiencies in dataset construction in this field. Commonly used CMFD datasets, such as CoMoFoD\cite{tralic2013comofod}, MMTDSet\cite{xu2024fakeshield} and COVERAGE\cite{wen2016coverage}, are based on natural images. Due to the significant differences in perspective and feature forms between remote sensing images and natural images, these datasets do not effectively support CMFD in remote sensing images. Currently, there are very few publicly available CMFD datasets\cite{zhang2024copymove} for remote sensing, and the tampering methods are randomly generated, lacking concealment and spatial authenticity, negatively impacting model training and evaluation.

\section{Dataset Construction}

The Real-RSCM dataset, comprising 10,000 images with a resolution of 512x512 pixels, was meticulously selected from the WHU-Building\cite{ji2018fully}, IAILD\cite{maggiori2017can}, HRSC\cite{liu2017high}, LAISFO\cite{waqas2019isaid}, iSAID\cite{kaiser2017learning}, and LoveDA\cite{wang_2021_LoveDA} datasets, encompassing 27 regions globally. It includes 8,620 copy-move tampered images and 1,380 untampered images as negative samples. Six kinds of salient targets—vehicles, airplanes, ships, buildings, roads, and vegetation—were manually scaled, rotated, and copied to reasonable locations within the images. Each tampered image was associated with at least 14 questions and 50 possible answers, resulting in 173,780 image-question-answer triplets. The distribution of questions and answers is depicted in Figure \ref{fig:dataset}(a). These questions can be categorized into basic, independent, and related questions, with specific examples illustrated in Figure \ref{fig:dataset}(b). Compared to the previous RS-CMQA\cite{zhang2024copymove} dataset, the Real-RSCM dataset offers the following significant advantages:


\begin{figure*}[!t]
    \centering
    \includegraphics[width=0.96\textwidth]{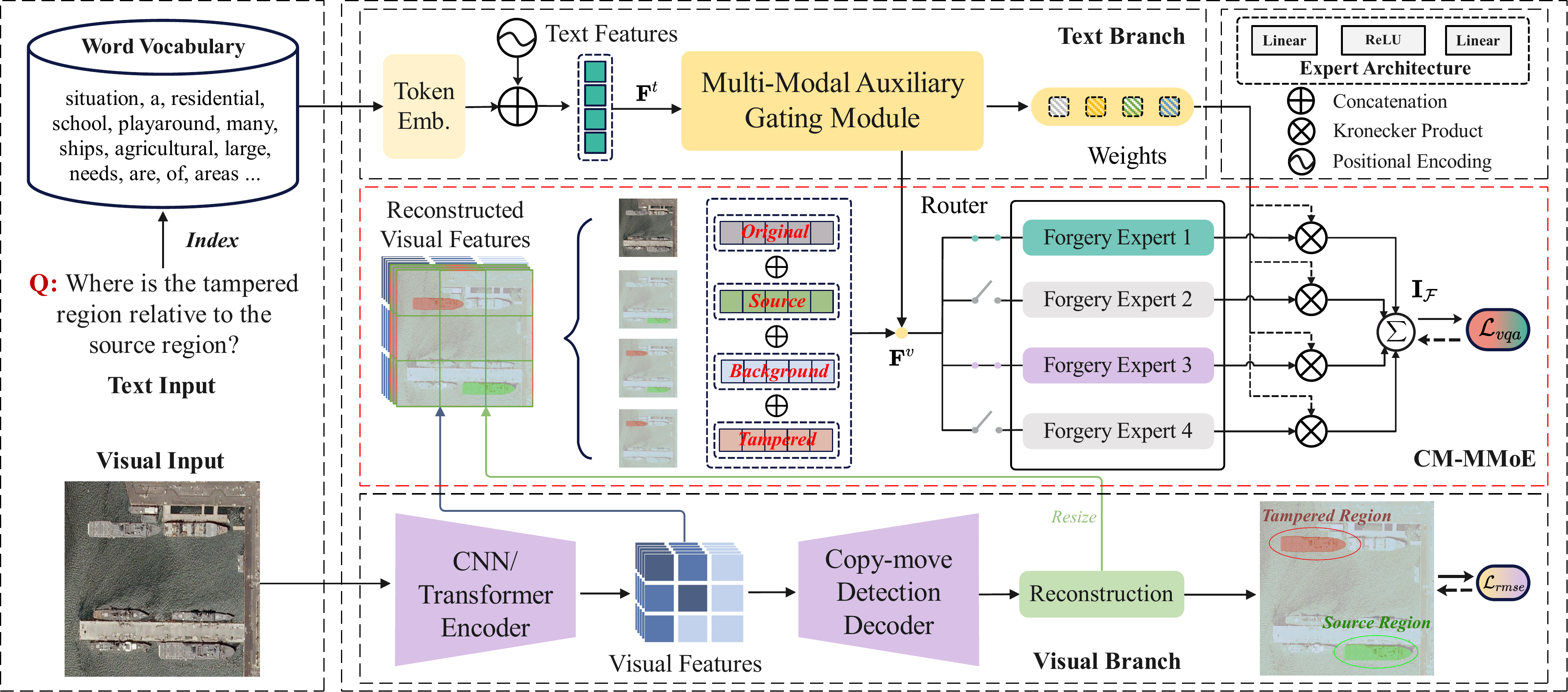}
    \vspace{-3pt}
    \caption{
    An illustration of the proposed CM-MMoE model.
    }
    \label{fig_framework}
    \vspace{-5pt}
\end{figure*}

a) The tampering in the Real-RSCM dataset is manually annotated, considering spatial rationality and concealment. This high-quality, challenging dataset includes difficult-to-detect tampered images, better simulating real tampering scenarios and providing more reliable model evaluations.

b) The Real-RSCM dataset features tampered objects with clear and distinct semantics, avoiding ambiguous annotations found in previous studies where fixed-wing aircraft and helicopters were considered the same, and vegetation and farmland were treated differently. This reduces noise and confusion, enhancing model training effectiveness.

\section{Methodology}

\subsection{Problem Formulation}

To identify the source region and tampered region and to perform precise reasoning by integrating foreground and background knowledge, a multimodal gated mixture of experts model (CM-MMoE) was designed. CM-MMoE involves a two-stage training process: 1) training the forgery detection network to generate hierarchical representations of the source region, tampered region, and background; and 2) utilizing the hierarchical multimodal representations for reasoning and answering. For the forgery detection network, masks of the source and tampered regions are used as ground truth to train the visual branch. The trained network outputs, along with the original image, serve as hierarchical representations for the VQA network.


\subsection{Forgery Detection for Visual Representation}
In scenarios with potential tampered regions, a novel reconstruction network is employed for visual representation.  Given an input image $\mathbf{I} \in \mathbb{R}^{H \times W \times 3}$, the encoder's output, which includes the background $\mathbf{F}^b$, source region $\mathbf{F}^s$, tampered region $\mathbf{F}^t$ masks, and the original input image $\mathbf{F}^o$, is utilized for hierarchical visual representation $\mathbf{F}^v \in \mathbb{R}^{H' \times W' \times C}$.  Here, $C$ represents the feature dimension.  The hierarchical visual modeling process is as follows:
\begin{equation}
{\mathbf{F}}^v=\operatorname{FFN}(\operatorname{Cat}(\mathbf{F}^o,\mathbf{F}^s,\mathbf{F}^b,\mathbf{F}^t)).
\end{equation}
Here, $\operatorname{FFN}$ denotes a feedforward neural network, and $\operatorname{Cat}$ represents the concatenation operation. Unlike element-wise sum and cross-attention, concatenation allows the background, source region, tampered region, and original image representations to retain their inherent characteristics without being disrupted by salient feature expressions. This facilitates the dynamic selection process in subsequent mixture of experts (MoE) models.

\subsection{Multimodal Mixture of Forgery Experts}
The main components of the MMoE include a multimodal auxiliary gating network $G_\mathcal{M}$ and a set of $N$ expert networks ${E_{1}^{forgery},...,E_{N}^{forgery}}$. In CM-MMoE, hierarchical visual representation $\mathbf{F}^v$ and textual representation $\mathbf{F}^t$ calculate similarity scores via a cross-attention mechanism. These scores serve as the basis for the multimodal auxiliary gating network to route the expert models. The corresponding gating weights for the $N$ expert networks are calculated as follows:
\begin{equation}
    G_\mathcal{M}(\mathbf{F}^v,\mathbf{F}^t) = Softmax(topK(Att_\mathcal{C}(\mathbf{F}^v,\mathbf{F}^t)\cdot W_\mathcal{M})),
\end{equation}
where $W_\mathcal{M}$ is a learnable weight matrix and the top $K$ outputs are normalized via $softmax$ distribution.

The structure of each expert network consists of $3$ FFN layers. Each expert takes the forgery multimodal fusion feature $\mathbf{F}^v$ as input to produce its own output $E_{i}^{forgery}(\mathbf{F}^v)$. The final output $\mathbf{I}_\mathcal{F}$ of CM-MMoE is the linearly weighted combination of each expert’s output with the corresponding gating weights. The formalization is as follows:


\begin{equation}
    \mathbf{I}_\mathcal{F} = \sum_{i=1}^{N}G_\mathcal{M}(\mathbf{F}^v,\mathbf{F}^t)_{i}E_{i}^{forgery}(\mathbf{F}^v).
\end{equation}

\begin{table*}[t]
\centering
\caption{
Evaluation with state-of-the-art methods on the Real-RSCM Test Set, with best metrics highlighted in \textbf{bold}.
}
\vspace{-3pt}
\resizebox{1\textwidth}{!}{
\renewcommand{\arraystretch}{1.1}
\large 
\begin{tabular}{c|ccc|cccccc|ccccc|cc}
\midrule[0.9pt]
 & \multicolumn{3}{c|}{Basic Questions} & \multicolumn{6}{c|}{Independent Questions} & \multicolumn{5}{c|}{Related Questions} &  &  \\
\multirow{-2}{*}{Method} & Q1 & Q2 & Q3 & Q4 & Q5 & Q6 & Q7 & Q8 & Q9 & Q10 & Q11 & Q12 & Q13 & Q14 & \multirow{-2}{*}{OA} & \multirow{-2}{*}{AA} \\ \midrule[0.9pt]
\multicolumn{1}{l|}{*General VQA Methods} & \multicolumn{1}{l}{} & \multicolumn{1}{l}{} & \multicolumn{1}{l|}{} & \multicolumn{1}{l}{} & \multicolumn{1}{l}{} & \multicolumn{1}{l}{} & \multicolumn{1}{l}{} & \multicolumn{1}{l}{} & \multicolumn{1}{l|}{} & \multicolumn{1}{l}{} & \multicolumn{1}{l}{} & \multicolumn{1}{l}{} & \multicolumn{1}{l}{} & \multicolumn{1}{l|}{} & \multicolumn{1}{l}{} & \multicolumn{1}{l}{} \\
SAN (CVPR, 2016)\cite{yang2016stacked} & 85.57 & 96.31 & 98.38 & 34.75 & 65.40 & 31.39 & 65.21 & 89.97 & 94.00 & 36.03 & 65.16 & 40.72 & 58.87 & 55.56 & 68.48 & 65.51 \\
MAC (ICLR, 2018)\cite{hudson2018compositional} & 87.94 & 98.62 & 99.15 & 49.54 & 84.27 & 64.10 & 85.56 & 94.06 & 95.28 & 82.12 & 91.12 & 77.74 & 88.16 & 55.05 & 84.80 & 82.34 \\
MCAN (CVPR, 2019)\cite{yu2019deep} & 84.44 & 96.39 & 98.75 & 26.18 & 68.19 & 35.15 & 75.23 & 90.13 & 92.99 & 34.86 & 73.26 & 71.43 & 61.74 & 54.60 & 71.84 & 68.81 \\
DVQA (NeurIPS, 2021)\cite{wen2021debiased} & 90.47 & 97.54 & 99.21 & 45.84 & 77.96 & 73.31 & 87.93 & 93.93 & 97.58 & 78.65 & 88.27 & 71.19 & 85.39 & 56.18 & 84.04 & 81.67 \\
BLIP-2 (ICML, 2023)\cite{li2023blip}  & \multicolumn{1}{l}{90.92} & \multicolumn{1}{l}{96.76} & \multicolumn{1}{l|}{99.17} & \multicolumn{1}{l}{48.24} & \multicolumn{1}{l}{74.58} & \multicolumn{1}{l}{48.32} & \multicolumn{1}{l}{78.52} & \multicolumn{1}{l}{88.50} & \multicolumn{1}{l|}{90.04} & \multicolumn{1}{l}{47.93} & \multicolumn{1}{l}{61.11} & \multicolumn{1}{l}{58.67} & \multicolumn{1}{l}{65.38} & \multicolumn{1}{l|}{58.36} & \multicolumn{1}{l}{74.04} & \multicolumn{1}{l}{71.89} \\ \midrule[0.9pt]
\multicolumn{1}{l|}{*Remote Sensing VQA Methods} & \multicolumn{1}{l}{} & \multicolumn{1}{l}{} & \multicolumn{1}{l|}{} & \multicolumn{1}{l}{} & \multicolumn{1}{l}{} & \multicolumn{1}{l}{} & \multicolumn{1}{l}{} & \multicolumn{1}{l}{} & \multicolumn{1}{l|}{} & \multicolumn{1}{l}{} & \multicolumn{1}{l}{} & \multicolumn{1}{l}{} & \multicolumn{1}{l}{} & \multicolumn{1}{l|}{} & \multicolumn{1}{l}{} & \multicolumn{1}{l}{} \\
RSVQA (TGRS, 2020)\cite{Sylvain2020RSVQA} & 88.63 & 96.29 & 98.25 & 48.19 & 80.98 & 60.66 & 84.86 &  92.54 & 95.73 & 74.04 & 86.52 & 42.91 & 61.60 & 54.46 & 78.77 & 76.12 \\
RSIVQA (TGRS, 2021)\cite{zheng2021mutual} & 84.16 & 95.18 & 98.18	 & 30.03 & 68.61 & 25.23	 & 68.39	& 91.22	 & 94.83 & 41.19 & 59.46 & 34.72 &  49.46 & 54.89 & 67.01 & 63.97 \\
FEH (TGRS, 2022)\cite{yuan2022easy} & 93.23 & 97.57 & 99.33 & 59.29	& 85.46 & 78.83 & 92.72 & 93.00	& 96.11 & 84.28 & 92.10 & 61.95 & 82.21 & 56.14 & 86.05 & 83.73 \\
MQVQA (TGRS, 2023)\cite{zhang2023multi} & 89.89 & 95.26 & 98.38 & 61.25 & 81.29 & 82.28 & 90.88 & 95.18 & 97.05 & 62.77 & 78.63 & 46.80 & 62.79 & 55.51 & 80.49 & 78.42\\
EarthVQA (AAAI, 2024)\cite{wang2024earthvqa} & 87.33 & 94.76 & 97.01 & 56.95 & 82.37 & 61.77 & 83.62 & 92.79 & 95.05 & 84.11 & 91.12 & 79.35 & 87.15 & 54.11 & 84.16 & 81.96 \\
SGA (IGARSS, 2024)\cite{tosato2024segmentation} & 89.40 & 95.83 & 96.97 & 57.45 & 83.33 & 63.01 & 85.08 & 94.79 & 97.24& 84.52 & 92.63 & 79.22 & 87.84 & 57.53 & 85.38 & 83.21\\
STMA (ArXiv, 2024)\cite{zhang2024copymove} & 91.79 & \textbf{97.78} & 99.05 & 68.82 & 87.13 & 82.28 & 92.25 & 97.52 & 97.90 & 79.79 & 90.08 & 68.85 & 82.78 & 55.86 & 87.08 & 85.12 \\ \midrule[0.9pt]
\rowcolor[HTML]{C8F3FF} 
\cellcolor[HTML]{C8F3FF}\textbf{CM-MMoE (Ours)} & \textbf{93.66} & 97.69 & \textbf{99.34} & \textbf{83.07} & \textbf{92.12} & \textbf{89.57} & \textbf{95.46} & \textbf{98.01} & \textbf{98.69} & \textbf{91.18} & \textbf{95.57} & \textbf{84.25} & \textbf{90.80} & \textbf{59.57} & \textbf{92.01} & \textbf{90.64} \\ \midrule[0.7pt]
\end{tabular}
}
\vspace{-7pt}
\label{tab:comparation}
\end{table*}

\subsection{Loss Function}
The loss function $\mathcal{L}$ is composed of tamper detection loss and VQA loss. The reconstruction loss for tamper detection is calculated based on Root Mean Squared Error (RMSE), while the VQA loss is determined using Cross Entropy (CE) loss. RMSE measures the discrepancy between the predicted source and tampered regions and the ground truth. Specifically, RMSE loss is given by:
\begin{equation}
\mathcal{L}_{\mathit{rmse}} = \sqrt{\frac{1}{n} \sum_{i=1}^{n} (\hat{\mathbf{F}}^o - \mathbf{F}^o)^2}
\end{equation}
where \( n \) denotes the number of samples, $\hat{\mathbf{F}}^o$ represents the ground truth mask, and $F^o$ is the predicted mask. The Cross-Entropy Loss for VQA is expressed as:
\begin{equation}
\mathcal{L}_{\mathit{vqa}} = -\frac{1}{n} \sum_{i=1}^{n} y_i \log(\hat{y}_i),
\end{equation}
where \( y_i \) denotes the ground truth answer and $\hat{y}_i$ represents the probability predicted through the fused representation $\mathbf{I}_\mathcal{F}$. Additionally, to balance the weights among experts, this work constructs a balanced loss based on the coefficient of variation. $\mathcal{L}_{\mathit{balance}}$ encourages all experts to have equal importance:
\begin{equation}
\mathcal{L}_{\mathit{balance}}=C V(\text {Importance}(\mathbf{F}^v,\mathbf{F}^t))^2
\end{equation}
Here, \text{Importance} represents $\sum_{x \in X} E(\mathbf{F}^v,\mathbf{F}^t)$, indicating the importance of the current expert in this batch of data. It is calculated as the sum of the gating network outputs corresponding to the expert for each sample in the batch. CV represents the coefficient of variation, determined by the ratio of the standard deviation to the mean. The loss $\mathcal{L}$ is defined as follows:
\begin{equation}
\mathcal{L} = \alpha \cdot \mathcal{L}_{\mathit{rmse}} + (1 - \alpha) \cdot \mathcal{L}_{\mathit{vqa}} + \mathcal{L}_{\mathit{balance}}  .
\end{equation}
where \(\alpha\) is a trade-off coefficient.

\section{Experiments}

\noindent \textbf{Evaluation metrics and Experimental settings.}
The overall accuracy (OA) for all questions and the average accuracy (AA) across all question categories were used to evaluate the model's predictive performance. Additionally, the accuracy for each question category was reported as a detailed reference metric. The CM-MMoE model was trained for 30 epochs, with the tampered mask reconstruction module updating weights only during the first 20 epochs. The learning rate for this module decays from 1e-3 to 1e-4, while the remaining weights decay from 5e-4 to 1e-6. Pretrained BERT and ViT-B were used as the feature encoder and the text feature encoder, respectively. Baseline models are configured according to parameters recommended in their respective original papers, with pretrained models used for encoders to ensure fairness. All experiments were conducted using the Pytorch framework on a single NVIDIA RTX4090 GPU. The Real-RSCM dataset was used for model training and evaluation, with 70\% of the data allocated to the training set, and 15\% each to the validation and test sets.

\subsection{Comparative Experiments}
Twelve advanced models were selected as baselines. These include SAN\cite{yang2016stacked}, MAC\cite{hudson2018compositional}, MCAN\cite{yu2019deep}, DVQA\cite{wen2021debiased} and BLIP-2-2.7B\cite{li2023blip} as classic general question-answering models, and RSVQA\cite{Sylvain2020RSVQA}, RSIVQA\cite{zheng2021mutual}, FEH\cite{yuan2022easy}, MQVQA\cite{zhang2023multi}, SGA\cite{tosato2024segmentation}, EarthVQA\cite{wang2024earthvqa}, and STMA\cite{zhang2024copymove} specifically designed for remote sensing tasks. The experimental results, as shown in Table \ref{tab:comparation}, indicate that all methods achieved satisfactory accuracy for basic questions. However, for independent and related Questions, which are strongly related to image forgery, baseline methods exhibited varying degrees of inaccuracy, highlighting the complexity and challenges of the RSCMQA task. SAN, MCAN, and RSIVQA attempted post-fusion feature enhancement with minimal success. Despite its large parameter count, BLIP-2-2.7B did not exhibit a performance advantage, suggesting that increasing model size without targeted feature extraction for tampered regions is ineffective. MAC and DVQA enhanced performance through specialized network designs and cross-modal feature alignment. FEH utilized question difficulty loss to train the model on challenging questions. EarthVQA and SGA extract semantic segmentation cues to assist in answering questions, resulting in some performance improvement. STMA specifically extracts information from the tampered and source regions, achieving relatively better performance. Our proposed CM-MMoE model achieved the best performance in 13 out of 14 question categories, with OA improving by 4.93\% and AA by 5.52\% compared to the second-best model. The CM-MMoE model extracts and aggregates various semantic information related to image tampering and employs the MMoE mechanism to flexibly understand comprehensive features at multiple levels according to different questions, providing accurate answers to a variety of questions. It is noteworthy that all models struggled with Question 14, which assesses whether an object has been rotated after copy-move tampering. This requires precise spatial localization of both the source and tampered regions, an area where current models still exhibit deficiencies.

\begin{table}[!t]
\centering
\caption{Results of Multimodal gating ablation experiment}
\vspace{-4pt}
\label{tab:mm}
\resizebox{0.95\linewidth}{!}{
\begin{tabular}{cc|ccccc}
\midrule[0.7pt]
Visual Feature & Text Feature & & OA & & AA  &\\ 
\midrule[0.7pt]
$\checkmark$ &  × & & 91.44 & &  89.82 & \\
× & $\checkmark$  &   & 90.83 & & 89.30 &  \\
$\checkmark$ & $\checkmark$ & & \textbf{92.01}   & & \textbf{90.64} & \\ 
\midrule[0.7pt]
\end{tabular}
}
\vspace{-4pt}
\end{table}

\begin{table}[!t]
\centering
\caption{Comparative experiments with other features fusion methods}
\vspace{-4pt}
\label{tab:fusion}
\resizebox{0.85\linewidth}{!}{
\renewcommand{\arraystretch}{}
\begin{tabular}{ccc|ccccc}
\midrule[0.7pt]
& Features Fusion& & & OA & & AA &\\ 
\midrule[0.7pt]
& Concatenate& & & 88.37 &   & 86.68  &\\ 
& CrossAttention  & & &  90.79 &  & 89.25 &\\
& Q-Former  & & &  90.93 &  &  89.39 &\\
\midrule[0.7pt]
& \textbf{MMoE(ours)} & & &  \textbf{92.01} & & \textbf{90.64} &\\ 
\midrule[0.7pt]
\end{tabular}
}
\vspace{-4pt}
\end{table}

\subsection{Ablation Experiments}

\noindent \textbf{Multimodal Gating Ablation.}
The Multimodal Gating module is designed to select the appropriate expert from multiple candidate expert modules. We attempted to use image features and text features independently as the basis for expert selection. As shown in Table \ref{tab:mm}, this approach led to a noticeable decline in performance, particularly when only text features were used. The experimental results indicate that guiding expert selection with multimodal features is both necessary and significantly advantageous.

\noindent \textbf{Features Fusion Ablation.}
Three classic feature fusion methods were compared with the MMoE approach, as shown in Table \ref{tab:fusion}. Feature concatenation lacks multimodal alignment and correlation. Although CrossAttention and Q-former improve feature alignment, they do not possess the flexibility of MMoE in addressing diverse question types and avoiding information redundancy. MMoE demonstrates a consistent performance advantage over these classic feature fusion methods.

\begin{figure}[!t]
    \centering
    \includegraphics[width=0.99\linewidth]{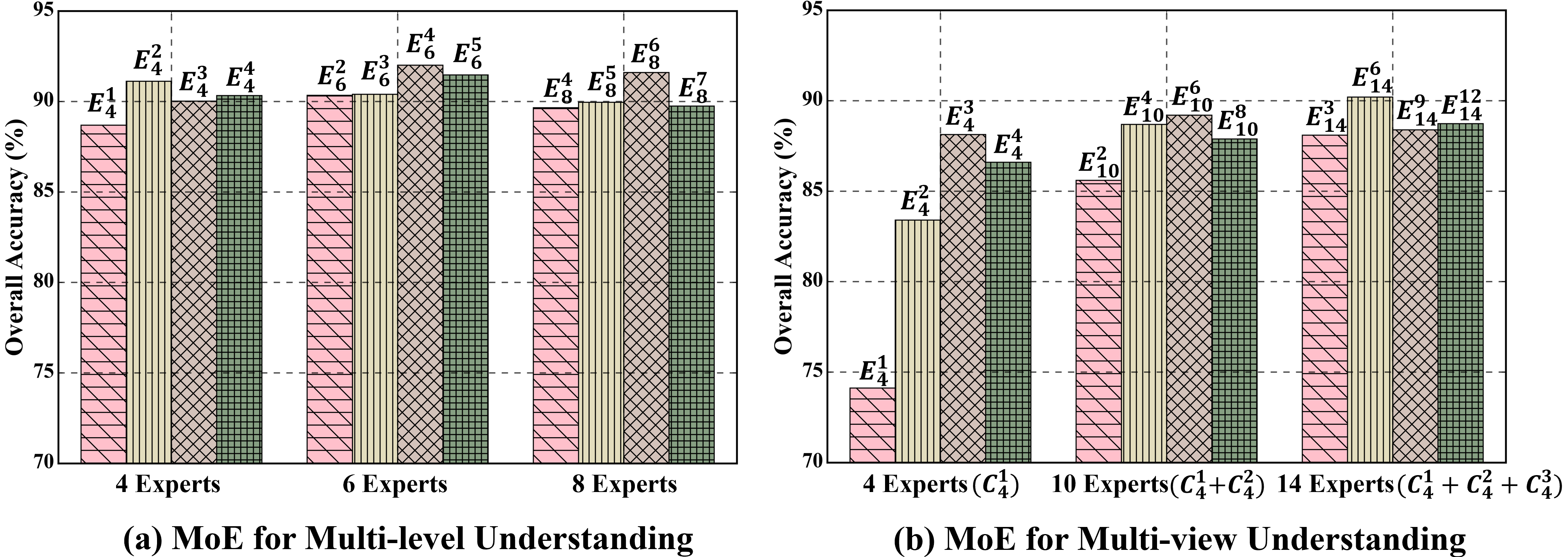}
    \vspace{-15pt}
    \caption{Examples of Expert structure ablation, where ${\mathbf{E}}_{m}^{n}$ representing $m$ experts and the top $n$ selected by multimodal gating. Multi-level understanding achieves better results with fewer experts than multi-view understanding. 
    }
    \vspace{-3pt}
    \label{fig:xiaorong}
\end{figure}

\noindent \textbf{Expert structure ablation.}
Extensive experiments were conducted on the structure and number of experts. Two design approaches for expert structure were evaluated. The first aggregates four types of features (source region, tampered region, background, and original image) into a joint feature ${\mathbf{F}}^v$, which different experts interpret in various ways, referred to as multi-level understanding. The second approach arranges the four feature types into different combinations, where each expert receives distinct features, such as ${\mathbf{C}}_{4}^{2}$ representing six combinations where each expert selects two features from the four, resulting in six experts. This is termed multi-view understanding.
The results, shown in Figure \ref{fig:xiaorong}, 
indicate that selecting too few top experts leads to insufficient information extraction, while selecting too many results in information redundancy. Multi-level understanding demonstrates a clear performance advantage over multi-view understanding, as each expert in the former has a comprehensive receptive field, providing more thorough information extraction. Conversely, multi-view understanding, despite more flexible feature selection, suffers from the limited receptive field's negative impact. The experiments show that with multi-level understanding, six experts, and selecting the top four through gating, the model achieves an optimal balance between feature extraction capability and avoiding information redundancy, yielding the best performance.

\begin{table}[!t]
\centering
\caption{Ablation experiment of loss function hyperparameter $\alpha$}
\vspace{-4pt}
\label{tab:hyper}
\resizebox{1\linewidth}{!}{
\begin{tabular}{c|ccccccc}
\midrule[0.8pt]
$\alpha$ & 0.1 & 0.2 & 0.3 & 0.4 & 0.5 & 0.6 & 0.7 \\ 
\midrule[0.8pt]
OA & 87.62 & 91.23 & \textbf{92.01} & 91.86 & 91.71 & 90.41  & 90.08 \\
AA & 86.05 & 89.89 & \textbf{90.64} & 90.55 & 90.30 & 88.96  & 88.47 \\
\midrule[0.8pt]
\end{tabular}
}
\vspace{-2pt}
\end{table}

\noindent \textbf{Loss Function Ablation.} The loss functions during the training process comprised $\mathcal{L}_{\mathit{rmse}}$ for detecting tampered regions, $\mathcal{L}_{\mathit{vqa}}$ for enhancing question-answering accuracy, and $\mathcal{L}_{\mathit{blance}}$ to weigh the importance among multiple experts. The $\mathcal{L}_{\mathit{blance}}$ quickly decreased and stabilized after training began, so no coefficient was applied to it. The sole loss hyperparameter, $\alpha$, was used to balance the importance of $\mathcal{L}_{\mathit{vqa}}$ and $\mathcal{L}_{\mathit{rmse}}$. As shown in Table \ref{tab:hyper}, the model achieved the best evaluation results for both OA and AA when $\alpha$ was set to 0.3.

\subsection{Visualization}

\begin{figure}[!t]
    \centering
    \includegraphics[width=\linewidth]{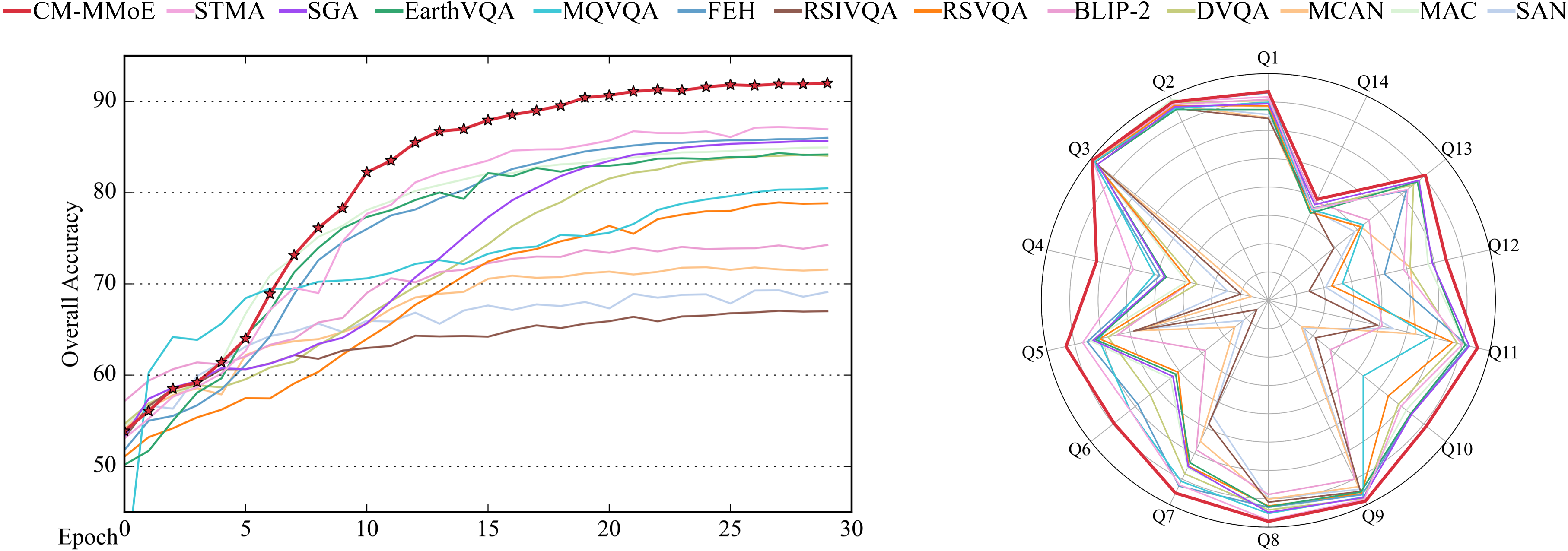}
    \vspace{-15pt}
    \caption{Left: Overall accuracy of the models per epoch on the validation set. Right: Accuracy coverage across various question categories on the test set. CM-MMoE demonstrates a stable and significant performance advantage.
    }
    \label{fig:vis}
\end{figure}

\begin{figure}[!t]
    \centering
    \includegraphics[width=\linewidth]{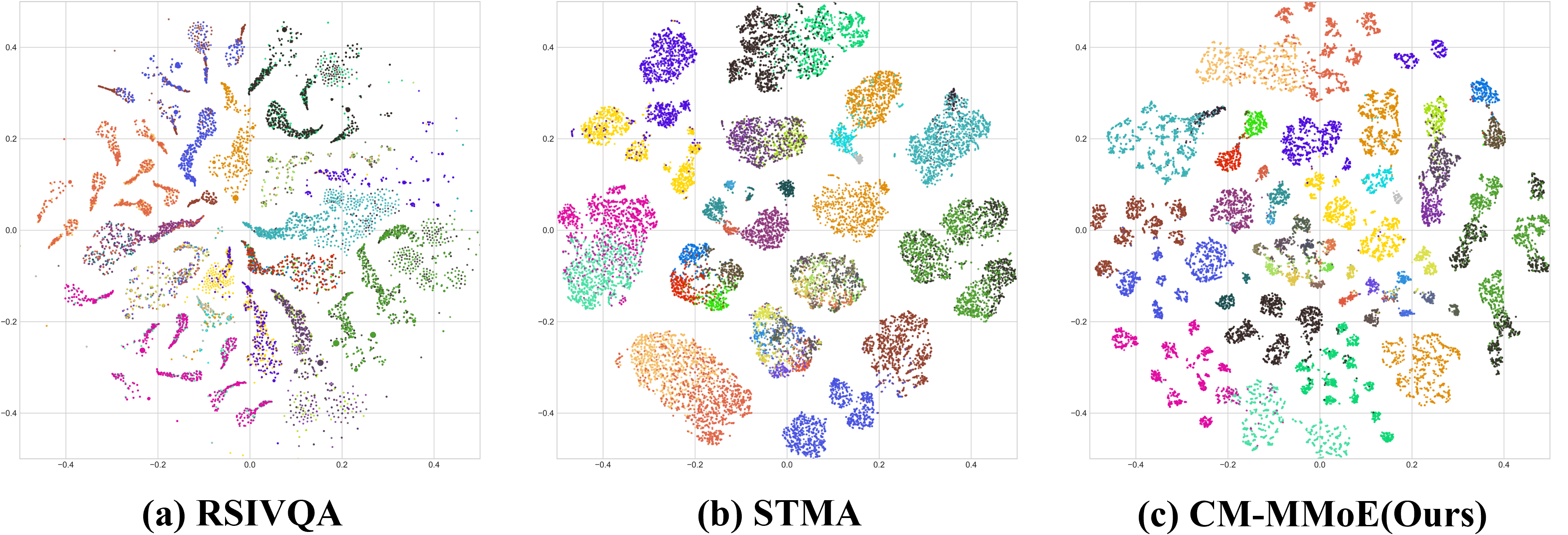}
    \vspace{-15pt}
    \caption{Dimensionality reduction visualization of feature vectors using t-SNE. RSIVQA exhibits a chaotic data distribution. STMA fails to effectively differentiate between answers within the same question category. In contrast, CM-MMoE demonstrates clear distinctions both between and within classes.
    }
    \label{fig:tsne}
\end{figure}

Figure \ref{fig:vis} illustrates the validation accuracy of CM-MMoE and baseline models during the training process, highlighting CM-MMoE's stable convergence trend and significant performance advantage. The accuracy coverage across various question categories on the test set is also shown in Figure \ref{fig:vis}, where CM-MMoE demonstrates a notable superiority over other baseline models when addressing complex questions. 
The feature vectors before the final linear layer of the model intuitively reflect the model's information extraction and aggregation effects. We performed dimensionality reduction visualization on these vectors using t-SNE. As shown in Figure \ref{fig:tsne}, compared to two classic baseline models, CM-MMoE exhibits clear data distribution for almost all answers, further proving its accurate discrimination capability for various complex questions.

\section{Conclusion}
In this study, we introduced Real-RSCM, a manually annotated, real, high-quality dataset for the RSCMQA task. We also proposed the CM-MMoE, a multimodal gated mixture of experts model, which can accurately answer diverse questions, providing a reliable benchmark for question answering in remote sensing tampering scenarios. Extensive experiments validated the superiority of the CM-MMoE model. Future work will focus on enriching the dataset with more diverse image tampering methods and question types. Additionally, we plan to explore integrating tampered region information into large multimodal models, advancing the application of remote sensing image tampering perception models in real-world scenarios.

\bibliographystyle{IEEEbib}
\bibliography{icme2025_template}


\end{document}